\documentclass[pra,superscriptaddress, twocolumn, longbibliography, reprint]{revtex4-1}

\usepackage{graphicx}
\usepackage{color}
\usepackage{amssymb}
\usepackage{amsmath}
\usepackage{mathtools}
\usepackage{bbm}
\usepackage{subfigure}
\usepackage{physics}
\usepackage[outdir=./figures/]{epstopdf}
\usepackage{soul}

\newcommand{\ceil}[1]{\left\lceil #1 \right\rceil}
\renewcommand{\imath}{\mathrm{i}}
\renewcommand{\vec}[1]{\mathbf{#1}}

\usepackage{xcolor}

\definecolor{smoothred}{HTML}{C5232F}
\definecolor{mygreen}{rgb}{0,0.5,0}
\definecolor{myblue}{rgb}{0,0,0.75}
\definecolor{mymagenta}{cmyk}{0,1,0,0.12}

\newcommand\redout{\bgroup\markoverwith
{\textcolor{red}{\rule[.5ex]{2pt}{0.4pt}}}\ULon}
\begin{document}

\title{Entanglement in the Quantum Game of Life}

\author{Peter-Maximilian Ney}
\affiliation{Theoretical Physics, Saarland University, D-66123 Saarbr\"ucken, Germany}

\author{Simone Notarnicola}
\affiliation{Dipartimento di Fisica e Astronomia ``G. Galilei'', via Marzolo 8, I-35131, Padova, Italy}
\affiliation{Padua Quantum Technologies Research Center, Universit\`a degli Studi di Padova.}
\affiliation{INFN, Sezione di Padova, via Marzolo 8, I-35131, Padova, Italy}

\author{Simone Montangero}
\affiliation{Dipartimento di Fisica e Astronomia ``G. Galilei'', via Marzolo 8, I-35131, Padova, Italy}
\affiliation{Padua Quantum Technologies Research Center, Universit\`a degli Studi di Padova.}
\affiliation{INFN, Sezione di Padova, via Marzolo 8, I-35131, Padova, Italy}

\author{Giovanna Morigi}
\affiliation{Theoretical Physics, Saarland University, D-66123 Saarbr\"ucken, Germany}

\date{\today}

\begin{abstract} 
We investigate the quantum dynamics of a spin chain that implements a quantum analog of Conway's game of life. We solve 
the time-dependent Schr\"odinger equation starting with initial separable 
states and analyse the evolution of quantum correlations across the lattice. We report examples of evolutions leading to all-entangled chains and/or to time oscillating entangling structures and characterize them by means of entanglement and network measures. The quantum patterns result to be quite different from the classical ones, even in the dynamics of 
local observables. A peculiar instance is a structure behaving as the quantum analog of a blinker, but that has no classical counterpart. 
\end{abstract}

\maketitle

\section{Introduction}

The Game of Life (GoL) is a celebrated cellular automaton originally proposed by Conway and since then used as a paradigm to study selforganization~\cite{gardner}. Its dynamics takes place on a network according to simple rules, it depends on the initial state and it is Turing-complete~\cite{Schulman1978,Rendell2002}. The implementation of the GoL in the quantum world has been object of increasing interest as a possible 
architecture for quantum computation~\cite{Arrighi}, for variational quantum optimization protocols and quantum state engineering~\cite{Brennen2003,Whittlock2020}. Moreover, it can be used to investigate the role of quantum mechanics in the emergence of physical complexity~\cite{RevModPhys.55.601,CarrMontangero}.
A central issue of quantum cellular automata is the characterization of the spreading of quantum information and of entanglement.
Recent works analysed the spreading of entanglement generated by quantum cellular automata~\cite{Brennen2003,CarrMontangero}, in
Ref. \cite{Piroli2020} an explicit relation is identified between quantum cellular automata and tensor networks, showing that the entanglement entropy satisfies an area law bound that can be generated at each dynamical step of the quantum cellular automata.

In this work we analyse the buildup of quantum correlations in the quantum game of life introduced in Ref.~\cite{EPL}: a Hamiltonian on a one-dimensional lattice simulating a reversible cellular automaton implementing the classical $F_{12}$-rule{, which swaps the state of a given site whenever either two or three sites of the nearest or next-nearest neihbours are alive~\cite{RevModPhys.55.601}}. We consider initial {separable} states with a well defined classical analog and determine their evolution by integrating the Schr\"odinger equation, both by methods based on ordinary differential equations as well as by means 
of tensor networks simulations~\cite{schollwoeck,Montangero2018a}. We compare the classical and the quantum dynamics obtained by evolving the same 
initial uncorrelated state and analyse the corresponding quantum correlations by means of quantum-information theoretical measures.  We show that the emerging quantum patterns exhibit qualitatively different characteristics from 
their classical counterpart in the behaviour of both local and non-local observables. One peculiar structure is the quantum blinker, namely a quantum recurrent structure that in the quantum GOL has no classical counterpart.

This manuscript is organized as follows. In Sec.\ \ref{Sec:model} we introduce the model and the observables we use in order to characterize the dynamics. In Sec.\ \ref{Sec:results} we present the results of the numerical integration of the Schr\"odinger equation for different initial states. An analytical model for the semiclassical analog of classical recurrent structures is derived 
in Sec.\ \ref{Sec:semi}. Finally, in Sec.\ \ref{Sec:conclusions} the conclusions are drawn and outlooks are discussed.

\section{A quantum game of life}
\label{Sec:model}

We consider a chain of $L$ quantum spins $1/2$. We denote the states of the spin $j$ by $\{\ket{0}_j, \ket{1}_j\}$, which are here chosen to be the eigenstates of the Pauli operator $\hat\sigma_z^j$.  We denote by $\hat 
b_j$ and $\hat b_j^\dagger$ the lowering and rising operators, which act on the basis states according to the relations $\hat b_j\ket{0}=0$, $\hat b_j^\dagger \ket{0}=\ket{1}$, and $\hat b_j^\dagger\ket{1}=0$, $\hat b_j \ket{1}=\ket{0}$, and such that $\hat\sigma_z^j=\hat b_j\hat b_j^\dagger-\hat b_j^\dagger\hat b_j$. We note that an eigenstate of the operator $\sum_{j=1}^L\hat\sigma_z^j$ is a sequence of $L$ bits. \\
\indent In what follows, $\ket{0}_j$ symbolizes the dead and $\ket{1}_j$ the alive cell, respectively. Let $\ket{\psi}_0$ be the initial state of 
the chain. Its evolution is governed by the Schr\"odinger equation 
\begin{equation}
\label{eq:S}
{\rm i}\partial_t\ket{\psi}_t=\hat H\ket{\psi}_t\,,
\end{equation}
where $t$ is a continuous dimensionless variable ($\hbar=1$). The dynamics is unitary, the state at time $t$ is $\ket{\psi}_t=\hat U(t)\ket{\psi}_0$ where $\hat U(t)=\exp(-{\rm i}\hat H t)$ is the time evolution operator.
Following Ref. \cite{EPL}, the model Hamiltonian reads
\begin{align}
    \label{eq:hamiltonian}
    \hat{H}\coloneqq \sum_{i=3}^{L-2}  \hat{\mathcal{S}}_i \qty(\hat{\mathcal{N}}_i^{(2)} + \hat{\mathcal{N}}_i^{(3)})\,.
\end{align}
The evolution it describes realises a continuous-time, quantum mechanical 
analog of the classical $F_{12}$-rule, where the state 
of a cell is swapped if and only if two or three cells among the nearest and next-nearest neighbours are alive (namely, in state $\ket{1}$).
Indeed, $\hat{\mathcal{S}}_i = \hat{b}_i + \hat{b}_i^\dagger$ and the operators $\hat{\mathcal{N}}_i^{(2),(3)}$ {are defined to be different from zero over the set of states where} only two (three) of the four nearest and next-nearest neighboring cells about $i$ are alive. They  are the sums of tensor products of projection operators $\hat{\bar n}_j\equiv \ket{0}_j\bra{0}$ and $\hat n_\ell\equiv \ket{1}_\ell\bra{1}$ for the nearest and next-nearest neighbouring sites $j$ 
and $\ell$ of site $i$ with eigenvalues is $0,1$. They take the form
\begin{align*}
    \hat{\mathcal{N}}_i^{(2)} = \phantom{+} &\hat{\bar{n}}_{i-2}\hat{\bar{n}}_{i-1}\hat{n}_{i+1}\hat{n}_{i+2}
    + \hat{\bar{n}}_{i-2}\hat{n}_{i-1}\hat{\bar{n}}_{i+1}\hat{n}_{i+2}\\
    + &\hat{\bar{n}}_{i-2}\hat{n}_{i-1}\hat{n}_{i+1}\hat{\bar{n}}_{i+2}
    + \hat{n}_{i-2}\hat{\bar{n}}_{i-1}\hat{\bar{n}}_{i+1}\hat{n}_{i+2}\\
    + &\hat{n}_{i-2}\hat{\bar{n}}_{i-1}\hat{n}_{i+1}\hat{\bar{n}}_{i+2}
    + \hat{n}_{i-2}\hat{n}_{i-1}\hat{\bar{n}}_{i+1}\hat{\bar{n}}_{i+2}\\
    \hat{\mathcal{N}}_i^{(3)} = \phantom{+} &\hat{\bar{n}}_{i-2}\hat{n}_{i-1}\hat{n}_{i+1}\hat{n}_{i+2}
    + \hat{n}_{i-2}\hat{\bar{n}}_{i-1}\hat{n}_{i+1}\hat{n}_{i+2}\\
    + &\hat{n}_{i-2}\hat{n}_{i-1}\hat{\bar{n}}_{i+1}\hat{n}_{i+2}
    + \hat{n}_{i-2}\hat{n}_{i-1}\hat{n}_{i+1}\hat{\bar{n}}_{i+2}\,,
\end{align*}
their eigenvalues are $0,1$.

We integrate numerically the dynamics for chains of finite size and for sufficiently long times. The resulting states and configurations are significantly affected by the boundary conditions. Here, we implement open boundary conditions, such that the state of cells {$j=1,\,2$ and $j=L-1,\,L$} is constant, as visible from Hamiltonian \eqref{eq:hamiltonian}.
The initial states we consider are separable and are eigenstates of the operator $\sum_{j=1}^L\hat\sigma_z^j$, that we denote below as Fock states and have a well defined classical analog.

We remark that Eq. \eqref{eq:hamiltonian} describes a unitary, and thus reversible, continuous-time dynamics. The classical GOL, instead, evolves according to a discrete-time, irreversible map fixed by a rule (the $F_{12}-$rule in our case). Here, in order to compare the dynamics of the two models we compare the unitary evolution with the classical analog. {The classical dynamics can also be derived by considering a modified stroboscopic dynamics of the quantum system in which} we set the duration of the time intervals to $\pi/2$. In fact, this is the time needed for a single spin in the state $\ket{0}$ or $\ket{1}$ to be completely swapped under the action of the operator $\hat{\mathcal{S}}_i$. At the beginning of each time step, moreover, we iteratively measure and fix the values of the projectors $\hat{\mathcal{N}}_i^{(2),(3)}$  and consequently let all the sites freely evolve or remain in their initial state. By doing so, if the initial state is a Fock state, then  the state at each step is a Fock state, and the sequence is the same as the one generated by the classical $F_{12}$-rule. %In order to characterize the emergence of complexity due to the quantum dynamics we solely consider initial Fock states, thus states with a very well defined classical analog. 

\subsection{Local observables}

We first discuss local observables, corresponding to measurements of operators $\hat n_i$ and $\hat{\bar{n}}_i$.  The evolution of the distribution of the alive cells is given by the expectation value 
\begin{align}
    \label{eq:population}
    n_i(t)=\ev{\hat{n}_i}\,,
\end{align}
and by its discretized description, 
\begin{align}
	\label{eq:dpopulation}
	\mathcal{D}_i(t) &\coloneqq \ceil{n_i(t)-0.5}\,,
\end{align}
which is either $1$ or $0$ depending on whether $n_i(t)>0.5$ or $n_i(t)\le 0.5$, respectively. According to this discretized description, the fraction of alive sites (density) is described by the quantity
\begin{align}
	\label{eq:density}
	\rho(t)&\coloneqq L^{-1}\sum_i \mathcal{D}_i(t)\,.
\end{align}
The formation of clusters of alive cells with size $\ell$ (with $0\le \ell\le L$) is revealed by the clustering function
\begin{align}
\label{eq:clusterfunction}
C(\ell,t)&\coloneqq \smash{\sum_{i=1}^{L-\ell+1}}  \,\hphantom{\cdot}\Theta_P\qty(\mathcal{D}_{i-1}(t)=0)\\
\nonumber&\hphantom{\coloneqq \sum_{i=1}^{L-\ell+1}} \cdot\Theta_P (\mathcal{D}_{j}(t)=1\, \forall\,j\in\qty{i,\ldots,i+\ell-1})\\
\nonumber&\hphantom{\coloneqq \sum_{i=1}^{L-\ell+1}} \cdot\Theta_P\qty(\mathcal{D}_{i+\ell}(t)=0)\,,
\end{align}
where here $\Theta_P$ denotes the Heaviside function for propositions $\mathcal A$, such that $\Theta_P(\mathcal A)=0$ if $\mathcal A$ is false while $\Theta_P(\mathcal A)=1$ otherwise.
It is useful to consider also the clustering function for dead cells:
\begin{align}
\label{eq:nclusterfunction}
\bar{C}(\ell,t)&\coloneqq \smash{\sum_{i=1}^{L-\ell+1}}  \,\hphantom{\cdot}\Theta_P\qty(\mathcal{D}_{i-1}(t)=1)\\
\nonumber&\hphantom{\coloneqq \sum_{i=1}^{L-\ell+1}} \cdot\Theta_P (\mathcal{D}_{j}(t)=0\, \forall\,j\in\qty{i,\ldots,i+\ell-1})\\
\nonumber&\hphantom{\coloneqq \sum_{i=1}^{L-\ell+1}} \cdot\Theta_P\qty(\mathcal{D}_{i+\ell}(t)=1)\,.
\end{align}
The number of different (living) cluster sizes present in the system at time $t$ is quantified by the diversity \cite{Schulman1978}:
\begin{align}
\label{eq:diversity}
\Delta(t)&\coloneqq \sum_{\ell=1}^{L}\Theta_0(C(\ell,t))\,,
\end{align}
that is defined using the Heaviside function $\Theta_0(x)$
 \cite{Schulman1978}. Furthermore, we will also use the so-called improved diversity, which counts over the average of dead and alive cells,  
\begin{align}
\label{eq:idiversity}
\Delta^{\text{impr.}}(t)&\coloneqq \frac{1}{2}\sum_{\ell=1}^{L}\Theta_0(C(\ell,t))+\Theta_0(\bar{C}(\ell,t))\,.
\end{align}
The diversity is a measure of complexity, in the sense that its increase is associated with an increase of the complexity \cite{Grassberger1986,Gros2015}. 

\subsection{Quantum correlations}

Quantum (non-classical) correlations are studied by means of the entanglement entropy $S(t)$ obtained by partitioning the system into a subset $A$ of cells and the rest of the lattice, which we denote by $B$ \cite{CarrMontangero}:
\begin{align}
\label{eq:entanglemententropy}
S(t)&= -\Tr\{\hat{\rho}_{A} (t)\log_2\qty(\hat{\rho}_{A}(t))\}\,,
\end{align}
where  the density matrix of the subset $A$ is obtained by tracing out the degrees of freedom of $B$, namely, $\hat\rho_A(t)=\Tr_B\{\ket{\psi}_t\bra{\psi}\}$, and $\Tr_B$ denotes the trace over a basis of subspace $B$. 

We characterize quantum correlations by means of different partitions of the lattice: We determine the {single}-point entanglement entropy $S_i(t)$ when $A$ consists of the single site $i$. The two-point entanglement entropy $S_{i,j}(t)$ is found when the subsystem $A$ consists of two sites $i,\,j$, which are not necessarily neighbours \cite{vargas,prl_mi}. These two quantities allow us to determine the mutual information,
\begin{align}
\mathcal{I}_{i,j}(t) &\coloneqq \frac{1}{2}\qty(S_i(t)+S_j(t)-S_{i,j})(t) 
\qq{with} \mathcal{I}_{i,i} \equiv 0\,,
\end{align}
which quantifies correlations between two sites \cite{inftheory}. The matrix $\mathcal{I}$ can be interpreted as the adjacency matrix of a weighted network. 
{
In order to quantify the entanglement between a pair of sites, we also use the concurrence between sites $i$ and $j$ \cite{Horo2009}. Given the two-sites density matrix $\rho$ and its complex conjugate $\rho^*$ expressed in the $\sigma_z$ basis, one defines the matrix $\tilde{\rho}=(\sigma_y\otimes\sigma_y)\rho^*(\sigma_y\otimes\sigma_y)$, the concurrence is
\begin{equation}\label{eq:conc}
\mathcal{C}_{ij}:= \mathrm{max}\{0,\lambda_1-\lambda_2-\lambda_3-\lambda_4\}\,,
\end{equation}
where $\lambda_\mu$ are the eigenvalues of the matrix $R=\sqrt{\sqrt{\rho}\tilde{\rho}\sqrt{\rho}}$  put in descending order. In our analysis we will use the averaged concurrence $\mathcal{C}(d)$ obtained by averaging over all the pairs of sites with distance $d$. Moreover,
}%chiusura parte in rosso
following Ref.~\cite{CarrMontangero}, one can determine typical network measures, such as the density of the network, which gives the fraction of existing links (correlations) between two nodes (cells) over all possible links:
\begin{align}
\label{eq:midensity}
    D &\coloneqq  \frac{1}{L(L-1)}\sum_{i,j=1}^L \mathcal{I}_{i,j}\,.
\end{align}
Further insight can be extracted from the average disparity $Y$  and from 
the clustering coefficient $C$: The first  quantifies the similarity of the network to a backbone \cite{Serrano2009,CarrMontangero}, while 
the second is a measure of the transitivity in a  network \cite{newman2003structure}. They are defined respectively as
\begin{align}
\label{eq:midisparity}
    Y &\coloneqq L^{-1} \sum_{i=1}^L \frac{1}{\qty(\sum_{k=1}^L \mathcal{I}_{i,k})^2}\sum_{j=1}^L \mathcal{I}_{i,j}^2\, ,\\
    %\frac{\sum_{j=1}^L \mathcal{I}_{i,j}^2}{\qty(\sum_{j=1}^L \mathcal{I}_{i,j})^2}\,,\\
\label{eq:miclustering}
    C &\coloneqq \frac{\Tr\qty(\mathcal{I}^3)}{\sum_{i=1}^L\sum_{j=1,\,j\neq i}^L \qty[\mathcal{I}^2]_{i,j}}\,.
\end{align}
We note that density, clustering, and disparity are here calculated using 
the mutual information, they thus provide insight into the behaviour of entanglement across the system, namely, a network whose nodes are the spins and whose edges are dynamically determined by Eq. \eqref{eq:S}.
Finally, entanglement across the lattice is quantified by means of the bond entropy $S_{\rm Bond}(t,j)$, which is found from Eq.~\eqref{eq:entanglemententropy} for a bipartition of the system at the bond between sites $j$ and $j+1$.

\section{Quantum dynamics}
\label{Sec:results}

In this section we discuss the dynamics of three types of initial states: (i) spatially localized states, where the initial clusters have unit size, (ii)  a cluster of a dozen cells, and (iii) an arbitrary  sequence of 0 
and 1 with varying density of alive cells, which we denote by random Fock 
state. The dynamics of these three kinds of states have 
been first analysed in Ref. \cite{EPL}. Here, we extend those studies by reporting the corresponding dynamics of quantum correlations and thereby elucidating the emergence of complexity with respect to the classical game of life. Let us remind that all the initial states we consider are separable Fock states. Therefore, they can be represented as a sequence of 0 and 1 and have a well defined classical analog.

In the following, for lattices up to 20 sites we numerically integrate the Schr\"odinger equation, Eq. \eqref{eq:S}, by means of fourth-order ordinary differential equations solver (ODE). For larger lattice sizes we resort to {Matrix Product States} (MPS) \cite{schollwoeck,Silvi2019} and the 
{Time Dependent Variational Principle} (TDVP) \cite{TDVP} from the library \emph{OpenMPS} \cite{ompsdoc,OpenMPS1,OpenMPS2}.

\subsection{Quantum blinkers}

\begin{figure*}
 \includegraphics[width=17cm]{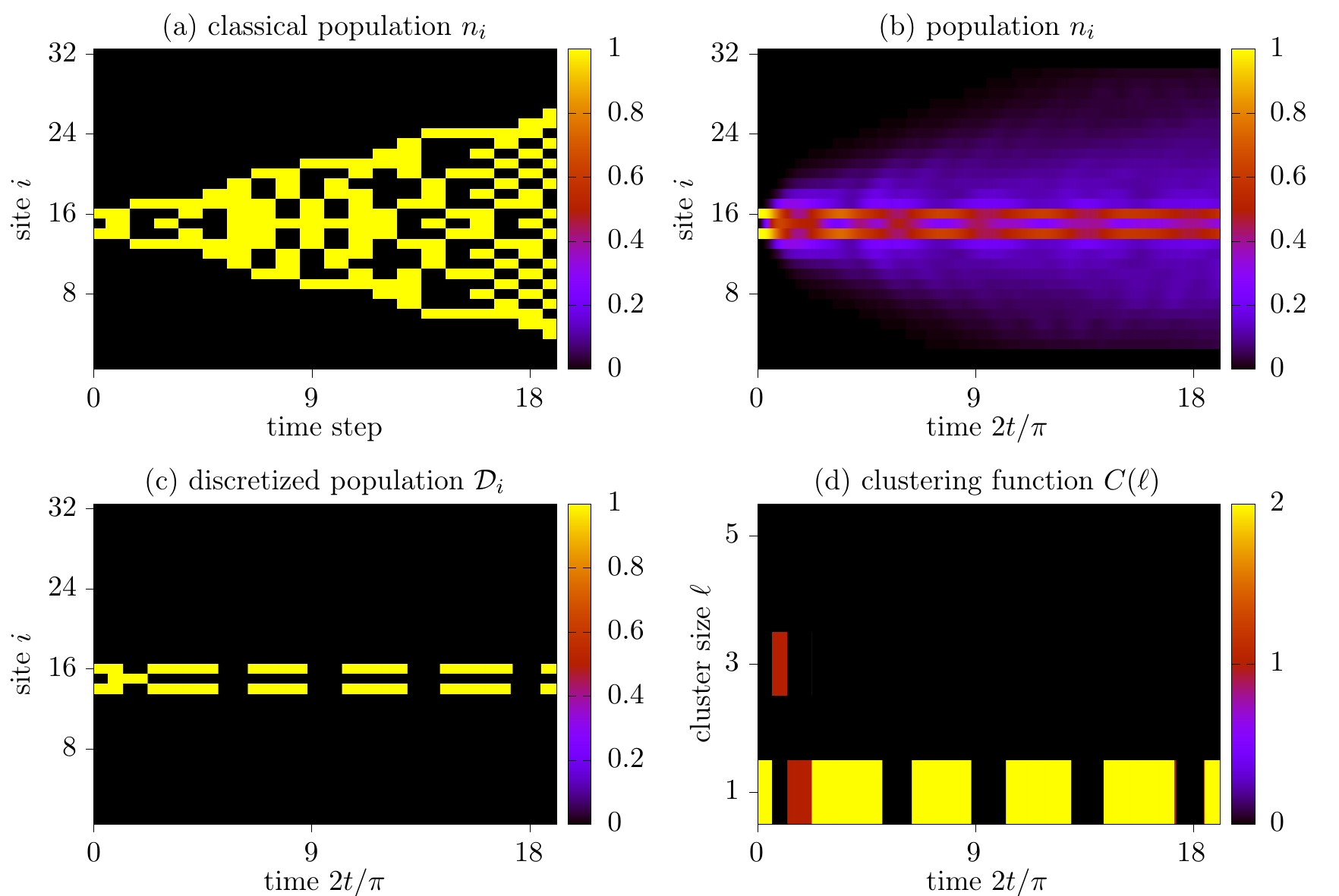}
    \caption{(color online) Time evolution of the initial state $\ket{\psi(0)}=\ket{0}^{\otimes 13}\otimes\ket{101}\otimes\ket{0}^{\otimes 16}$ (the quantum blinker) according to (a) the classical $F_{12}$ rule and (b,c,d) the Schr\"odinger equation, Eq. \eqref{eq:S}. Suplot (a) shows the alive cells as a function of time according to the classical GOL. Subplots (b) and (c) display respectively the expectation value $n_i$ over the quantum state, Eq. \eqref{eq:population}, and the corresponding discretized population $\mathcal{D}_i(t)$, Eq. \eqref{eq:dpopulation}. Subplot (d) shows the surface plot of the clustering function $\mathcal{C}_{\ell}(t)$, Eq. \eqref{eq:clusterfunction} as a function of time $t$ and cluster size $\ell$.  The time evolution is performed over a lattice of 
$L=32$ sites using TDVP with bond dimension $m=32$ and time step $\Delta t=0.01$. Note that in the classical dynamics (subplot (a)) the evolution occurs over a discrete time grid of step $\pi/2$.} 
    \label{fig:1}
\end{figure*}

\begin{figure*}
 \includegraphics[height=5cm]{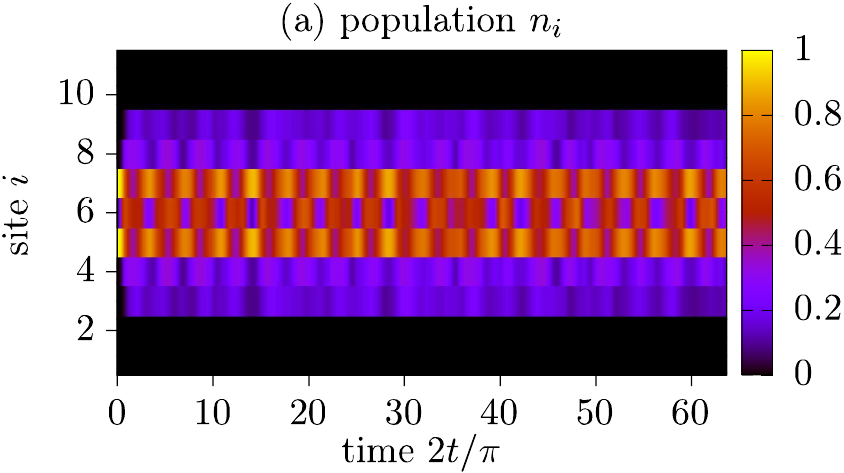}\, \,
 \includegraphics[height=5cm]{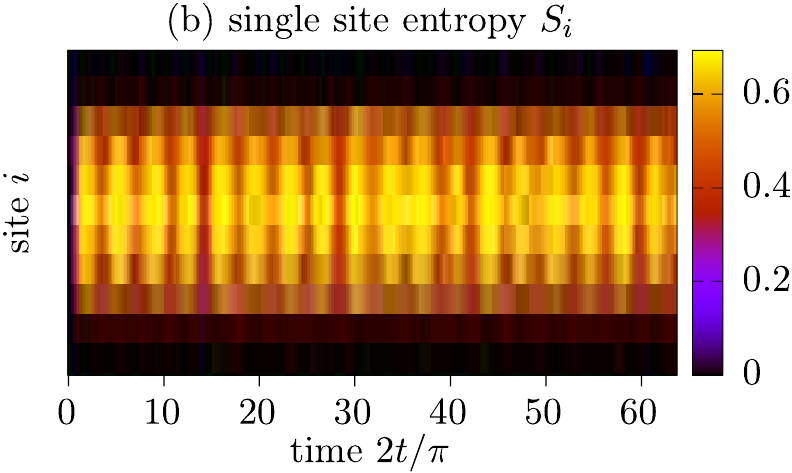} \\ \vskip 0.3cm
 \includegraphics[height=6.1cm]{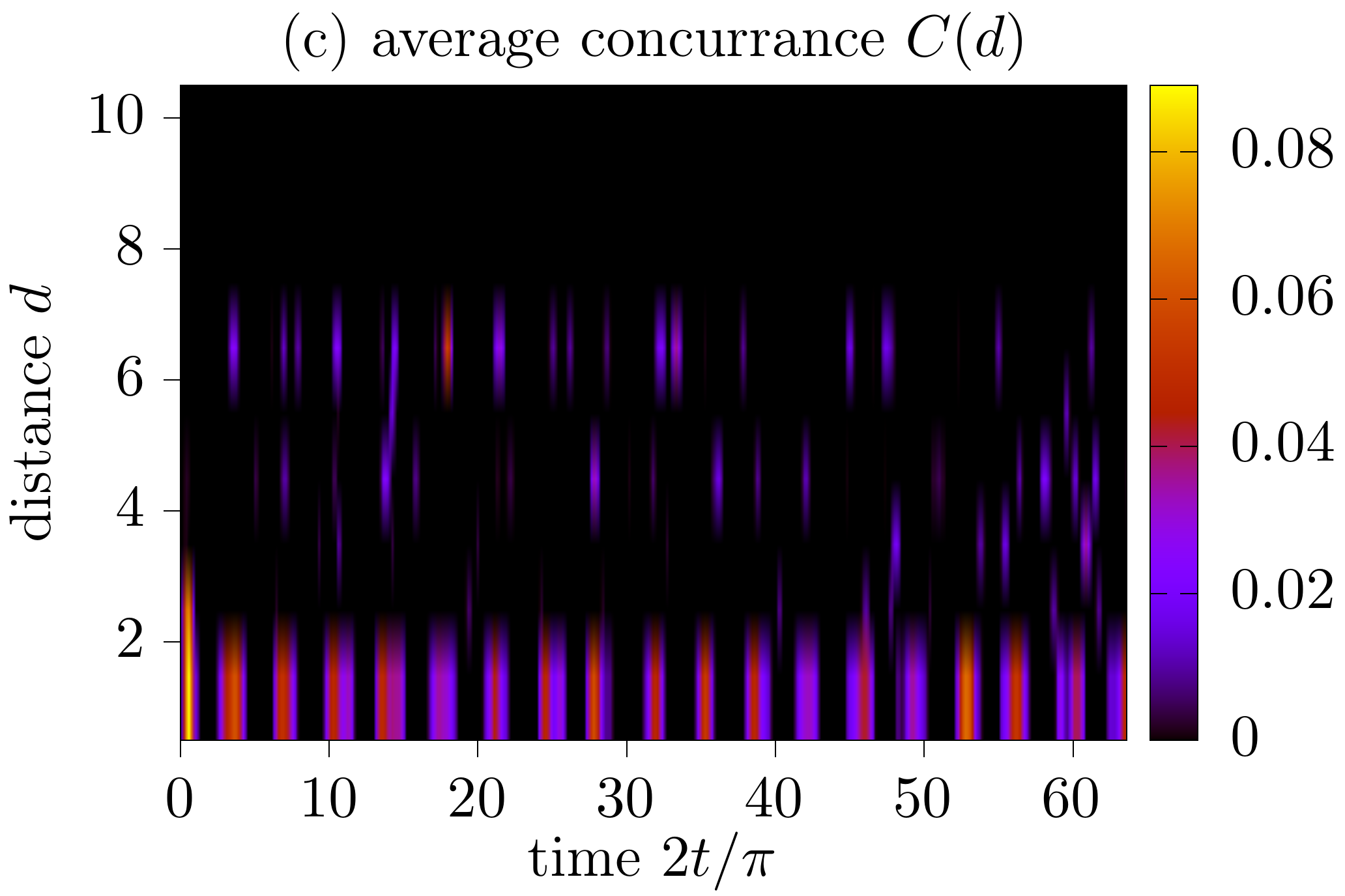} 
    \caption{{(color online) Time evolution of the initial state $\ket{\psi(0)}=\ket{0}^{\otimes 4}\otimes\ket{101}\otimes\ket{0}^{\otimes 4}$ (the quantum blinker) according to the Schr\"odinger equation, Eq. \eqref{eq:S}. Suplot (a) shows the expectation value $n_i$ over the quantum state, Eq. \eqref{eq:population}, and the corresponding (b) singe site entropy $S_i$. Panel (c) shows the average concurrence $C(d)$, Eq. \ref{eq:conc} The time evolution is performed over a lattice of 
$L=11$.}} 
    \label{fig:2}
\end{figure*}

We consider a spatially localized state with a string $\ket{1,0,1}$ embedded in a sequence of zeros. Figure \ref{fig:1}(a) displays its evolution for  the classical model, in which the distribution of alive cells spreads from the center of the lattice with an approximate diffusion rate of half lattice cell per time step.
The quantum dynamics, see subplot (b), is instead characterized by dynamical localization of the cell distributions about the initial positions of 
the alive cells. The behaviour becomes more manifest in the dynamics of the corresponding discretized population, visible in subplot (c). We here observe collapses and revivals of the alive cells. 
These features are well captured by the clustering function, (d), where the dynamics alternate no clusters with two clusters of alive cells and of unit size. This kind of dynamics has been first reported in Ref.~\cite{EPL} and has been called``quantum blinker'' for the recurrent behavior reminiscent of the blinkers of Conway's game of life. 
{It is possible to relate the dynamics of the local population $n_i$ with the single-site entropy. We consider a single blinker on a chain with $L=11$ and evolve it as shown in Fig. \ref{fig:2}. In Fig. \ref{fig:2}(a) we show the dynamics of the local occupation $n_i$ while in Fig. \ref{fig:2}(b) we plot its single-point entropy.
By looking, in particular, 
at the central site entropy one can notice an oscillating behavior. 
We note that the revivals of the initial structure of the blinker coincide in time with the minima in the single-site entropy, showing that the entanglement of the chain oscillates and it is minimal when the revivals occur. This behavior shares analogies with those due to dynamical symmetries in closed and open quantum systems. \cite{PhysRevB.102.041117,Buca2019}.
}
As visible from comparison with Fig. \ref{fig:1}(a), the recurrent features of this dynamics emerge only in the quantum regime, and are clearly a result of the interplay between interactions and quantum interference.  

\begin{figure*}
\includegraphics[width=17cm]{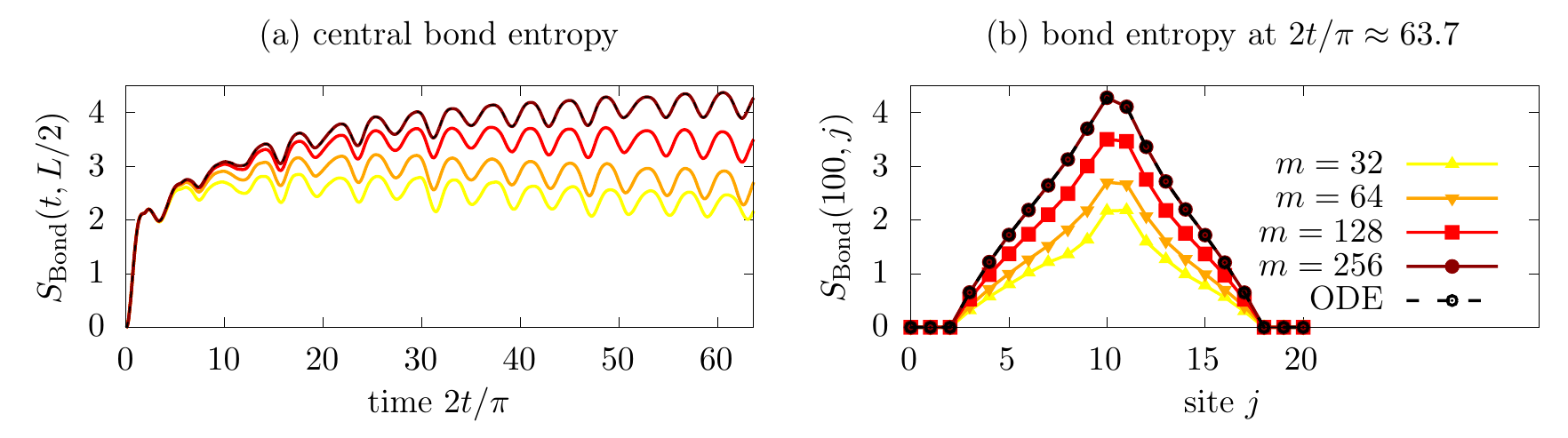}
    \caption{(color online) Quantum dynamics of correlations across the lattice with size $L=20$ and for the initial state  $\ket{\psi(0)}=\ket{0}^{\otimes 9}\otimes\ket{101}\otimes\ket{0}^{\otimes 8}$ (quantum blinker). (a) Central bond entropy $S_{\rm Bond}\qty(t,\frac{L}{2})$, Eq. \eqref{eq:entanglemententropy}, as a function of time and (b) bond entropy $S_{\rm Bond}(t,j)$ as a function of 
the cell $j$ and evaluated at time 100. The curves are determined using TDVP and bond dimensions $m=32,64,128,256$ {(darker colors correspond to higher bond dimension values, see legenda)}. The data linked by the dashed line correspond to a simulation using ordinary differential equations (ODE). The time step of the evolution is $\Delta t=0.01$.}
 \label{fig:3}
\end{figure*}

In order to study the spreading of nonlocal correlations we also compute the central bond entropy by taking a single blinker as initial state and simulating its dynamics at $L=20$. The corresponding dynamics is displayed in Fig. \ref{fig:3}(a). After a transient, the central bond entropy increases linearly and then starts performing oscillations about a slowly increasing mean 
value. These oscillations are out-of-phase with respect to the oscillations of the density pattern. 
Figure \ref{fig:3}(b) reports the bond entropy at $t=100$ and as a function of the lattice sites: the bond entropy is maximum when partitioning 
the chain at the center (which is here the center of the blinker) and decreases linearly when approaching the edges, suggesting 
that entanglement follows a volume-law at sufficiently long times. This conjecture is supported by the scaling of the central bond entropy, computed with tensor networks, as a function the bond dimension $m$: the entanglement values converge to the ones of the ODE only for bond dimension $m=256$ {(dark brown line)}, which corresponds to considering the full Hilbert space.
The growth of the entanglement and the emerging volume law are consistent 
with the results of Ref. \cite{Schachenmayer_2013} about quench dynamics in short ranged spin Hamiltonians, where it is shown 
that the entanglement grows linearly in time until a saturation value, proportional to the partition size, is reached. This saturation value is due to the counter propagation of entangled quasi particles along the lattice. The dynamics we report, moreover, can be compared with the scenario considered in Ref. \cite{Piroli2020}, in which a single quantum cellular automaton (QCA) application is 
discussed. In fact, we can cast our problem in terms of the dynamics of a continuous-time QCA over an infinitesimal time. In these terms our findings are consistent with the area-law entanglement bound of Ref. \cite{Piroli2020}.

\begin{figure*}
\centering
  \includegraphics[height=4.6cm]{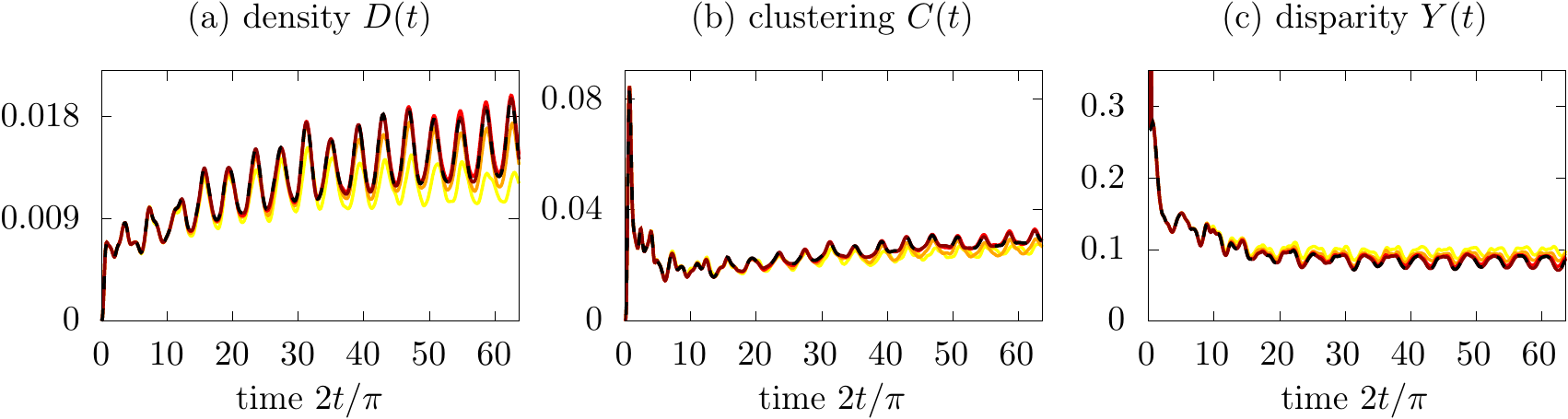}
    \caption{(color online)
Characterization of the dynamics of a quantum blinker using network measures. (a) Density $D(t)$, Eq. \eqref{eq:midensity}, (b) clustering $C(t)$, 
Eq. \eqref{eq:miclustering}, (c) disparity $Y(t)$, Eq.  \eqref{eq:midisparity}. The parameters and legenda are the same as in Fig. \ref{fig:3}. }
    \label{fig:2b}
\end{figure*}

{Network measures provide further information on complexity and insights on the entanglement properties of the system \cite{doggen2021generalized}}. Figure \ref{fig:2b}(a) displays the dynamics of the density of entangled spin pairs. The mean value of this quantity slowly grows. The oscillations have 
relatively large amplitude and show that collapses and revivals of the density are accompanied by oscillations of the quantum correlations. This is a 
further manifestation of the quantum dynamics underlying this phenomenon. 
These oscillations are also visible in the clustering (b) and in the disparity (c). The relative amplitude is visibly smaller. The behaviour of clustering reflects the one visible of the clustering function in Fig. \ref{fig:1}(d): the initial increase is associated  with the formation of a large cluster, that then decays into smaller clusters. The slow asymptotic growth of the clustering can be attributed to the spreading of correlations across the chain. The disparity, (c), seems instead to settle about a constant mean value, suggesting that the backbone of the network has reached a stationary structure about which it oscillates.
{
The behavior of the network measures can also be related to the dynamics of the average concurrence $\mathcal{C}(d)$ defined in Eq. (\ref{eq:conc}). Indeed, the oscillations in the concurrence at $d=1$, shown in Figure \ref{fig:2}(c) resemble those of the density and the clustering highlighted before. The fact that the concurrence vanishes while the bipartition bond entropy grows linearly in time, suggests a multipartite entanglement structure.
}

\begin{figure*}
    % GNUPLOT: LaTeX picture with Postscript
\includegraphics[height=4.8cm]{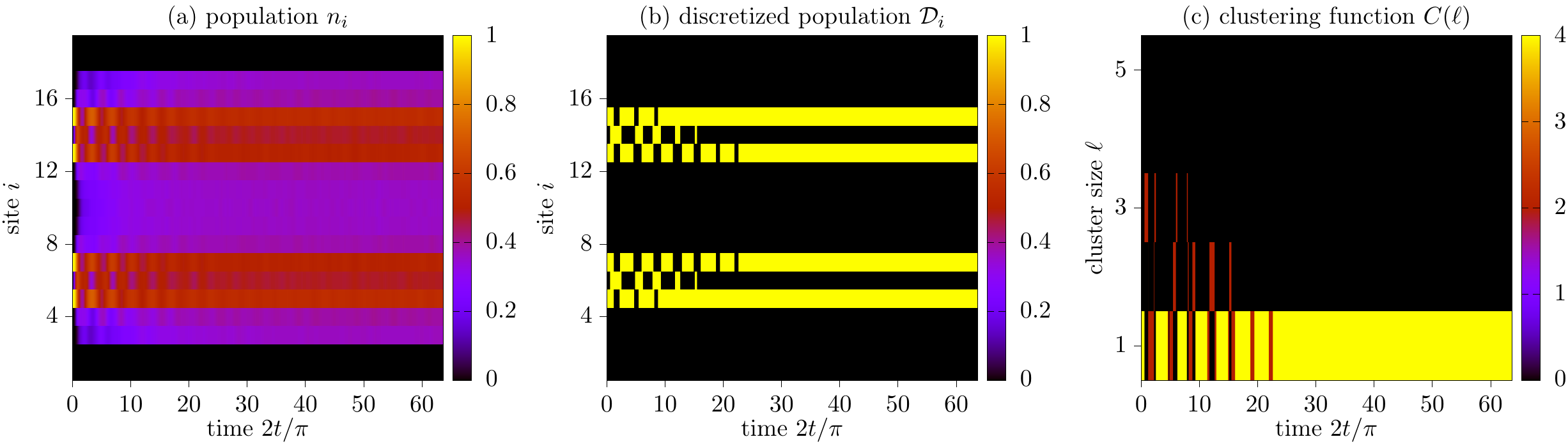}
    \caption{(color online)
        Dynamics of the interaction between two quantum blinkers. Subplot (a) is the surface plot of the density of each site as a function of time
for the dynamics of the double-quantum-blinker (``two quantum blinkers") state $\ket{\psi(0)}=\ket{0000}\otimes\ket{101}\otimes\ket{00000}\otimes\ket{101}\otimes\ket{0000}$  
and  (b)  the corresponding discretized population. Subplot (c) gives the  evolution of the clustering function for the two quantum blinkers. The size of the lattice is $L=19$. The time
step is $\Delta t=0.01$.}
    \label{fig:b2b}
\end{figure*}

\begin{figure*}
   \includegraphics[width=17cm]{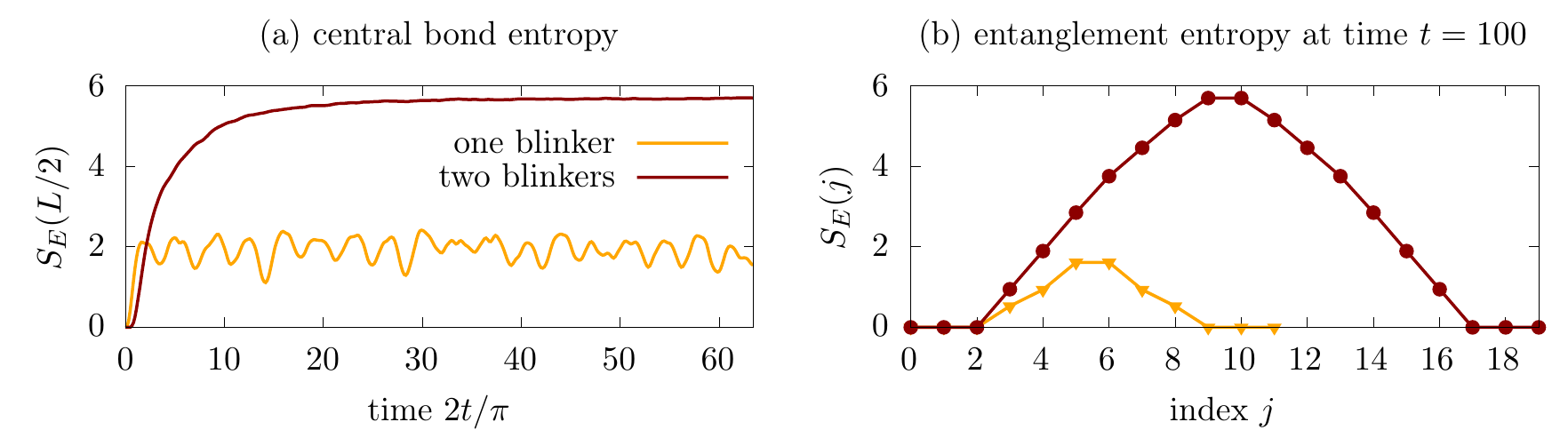}
    \caption{(color online)
      {Dynamics of the central bond entropy (a) and equilibrium bond entropy as a function of the lattice sites (b) for one (yellow, light curve, $L=11$) and two (brown, dark curve, $L=19$) quantum blinkers. The bond entropy for a single quantum blinker is shown for comparison.}
}
    \label{fig:b2b:entropy}
\end{figure*}

Quantum blinkers are reminiscent of discrete breathers \cite{Flach1998} and are a manifestation of quantum interference. We now report the numerical study of  the interaction of an initial state composed 
by two spatially separated clusters of cells $\ket{101}$. For infinite distances between the two clusters, we would observe two quantum blinkers. We plot their dynamics in Fig. \ref{fig:b2b} for a finite distance. In particular, Fig. \ref{fig:b2b}(a) shows the dynamics of the local occupation expectation value $n_i$. The surface plot of the density in Fig. \ref{fig:b2b}(b) shows that blinking structures interfere with each other. This effective interaction tends to destroy the collapse and revivals of the discretized population. Nevertheless, the discretized density, Fig. \ref{fig:b2b}(c), remains localized about the position of the individual blinkers. The stationary bond 
entropy, shown in Fig. \ref{fig:b2b:entropy}, exhibits a maximum when the 
bond is taken at half distance between the two structures, suggesting that the blinkers become entangled with one another. A similar behaviour has been reported for other Hamiltonian dynamics simulating quantum cellular automata \cite{CarrMontangero, Farrelly2020reviewofquantum, arrighi2019overview}. Further studies shall analyse the dependence of these features on the intermediate distance between the blinkers and on their number across the lattice.

\subsection{Evolution of a cluster of alive cells.}

\begin{figure*}
 \includegraphics[width=17cm]{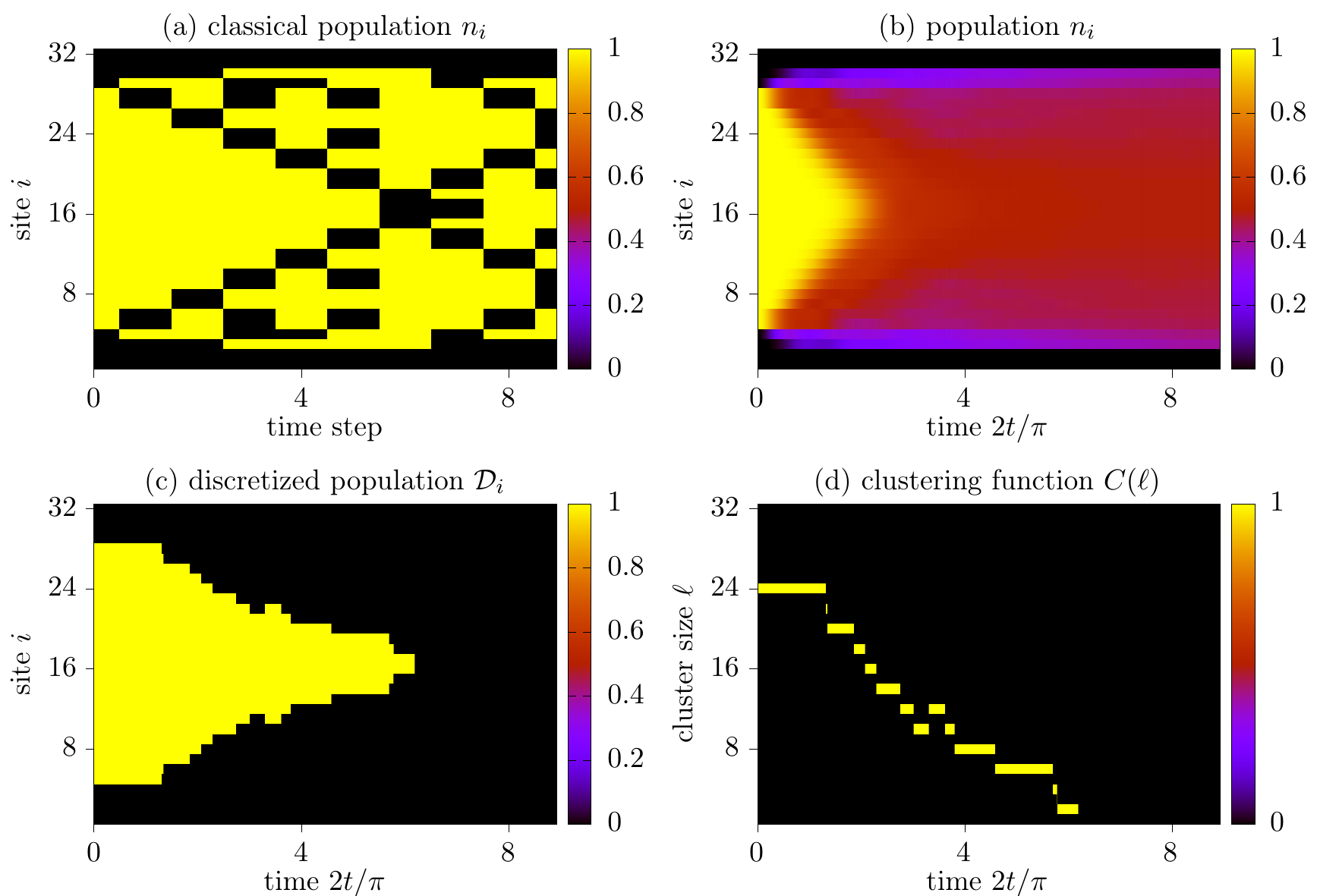}
    \caption{(color online) Characterization of the time evolution of the initial state $\ket{\psi(0)}=\ket{0}^{\otimes 4}\otimes\ket{1}^{\otimes 24}\otimes\ket{0}^{\otimes 4}$. The subplots and parameters are the same as Fig. \ref{fig:1}, except that here for the TVDP simulations the bond dimension is $m=256$.} 
    \label{Fig:cluster}
\end{figure*}

\begin{figure*}
  \includegraphics[width=17cm]{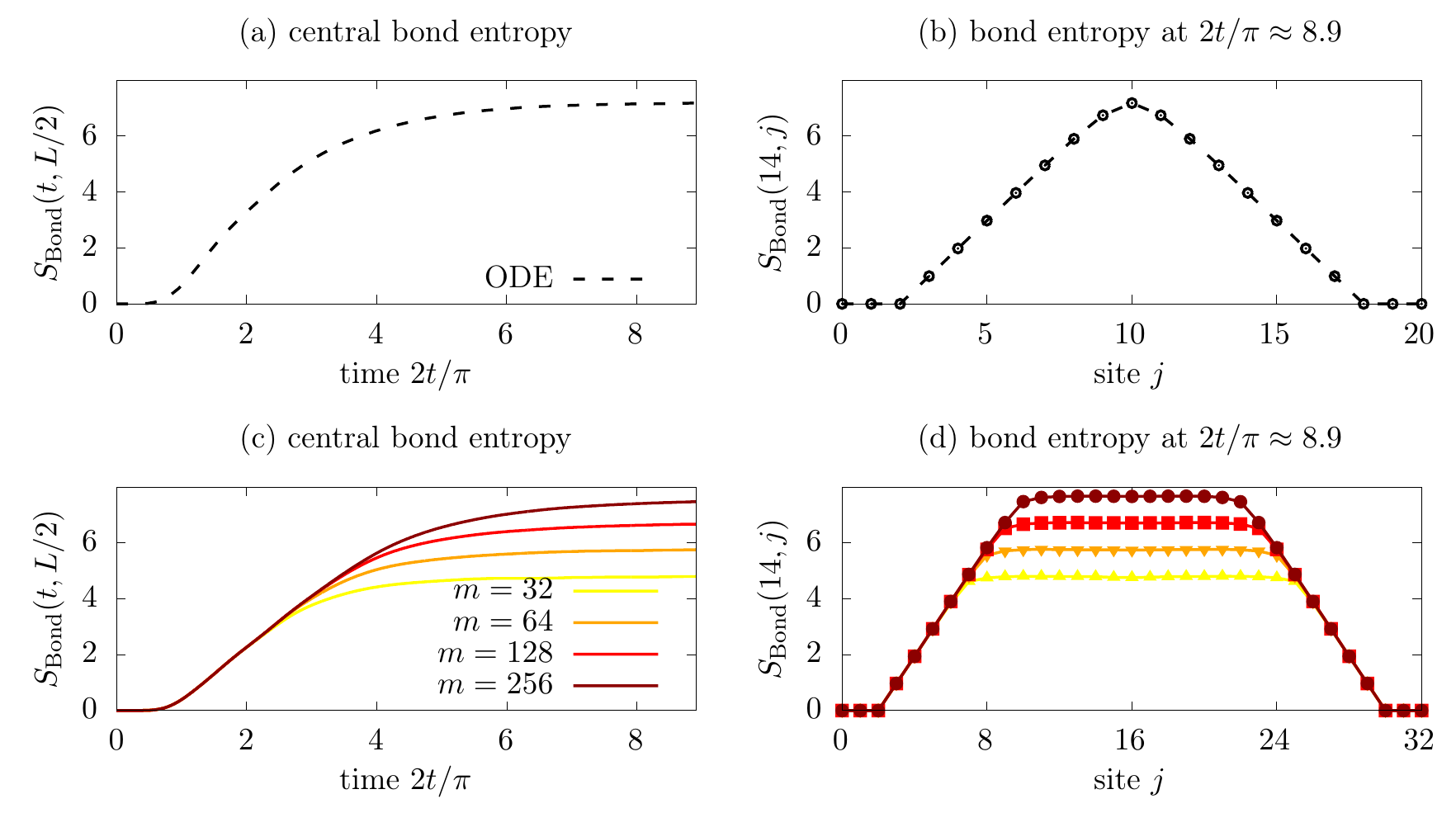}
    \caption{(color online)
    Quantum dynamics of correlations across the lattice with size $L=16$ (top row) and $L=32$ (bottom row) and for the initial state $\ket{\psi(0)}=\ket{0}^{\otimes 4}\otimes\ket{1}^{\otimes 12}\otimes\ket{0}^{\otimes 4}$ (top row) and $\ket{\psi(0)}=\ket{0}^{\otimes 4}\otimes\ket{1}^{\otimes 24}\otimes\ket{0}^{\otimes 4}$ (bottom row). (a-c) Central bond 
entropy $S_{\rm Bond}\qty(t,\frac{L}{2})$, Eq. \eqref{eq:entanglemententropy}, as a function of time and (b-d) bond entropy $S_{\rm Bond}(t,j)$ as a function of the cell $j$ . The curves are determined using TDVP and bond dimensions $m=32,64,128,256$ {(darker colors correspond to higher bond dimension  values, see legenda)}. The data linked by the dashed line correspond to a simulation using ordinary differential equations. The time step of the evolution is $\Delta t=0.01$.}
\label{fig:cluster:entropy}
\end{figure*}

We now consider a different initial state, composed by a single cluster of alive cells with size $\sim L$. The dynamics of the local density for the classical and quantum model are displayed in Fig. \ref{Fig:cluster}. The classical dynamics leads to the propagation of patterns of dead cells from the edges, that seem to follow a classical trajectory. This leads to a fragmentation of the initial cluster. After 6 time steps, where the two trajectories starting from the edges meet at the chain center, the initial cluster of alive cells is splitted into two large ones. On the contrary, in the quantum dynamics the size of the cluster decreases as a function of time, such that after a time evolution corresponding to 6$\pi/2$ the clustering function vanishes. On this time scale the density reaches a uniform value across the lattice which is below $1/2$, and thus smaller than the initial average density. The dynamics of the bond entropy is shown in Fig. \ref{fig:cluster:entropy}(a) for $L=16$ and  Fig. \ref{fig:cluster:entropy}(c) for $L=32$: Entanglement grows  across the lattice until reaching a saturation value. The spatial profile at long times is shown in Figs. \ref{fig:cluster:entropy}(b),(d) and displays a volume law, as for the quantum blinker. Notice that the entropy spatial profile has not converged for $L=32$ at bond dimension $m=256$ and  $t\geq 4\pi/2$. 
%\textcolor{red}{The bond entropy growth as a function of the subsystem size is consistent with a final state composed by a uniform superposition of all the states in the Hilbert space.}

\begin{figure*}
   \includegraphics[width=17cm]{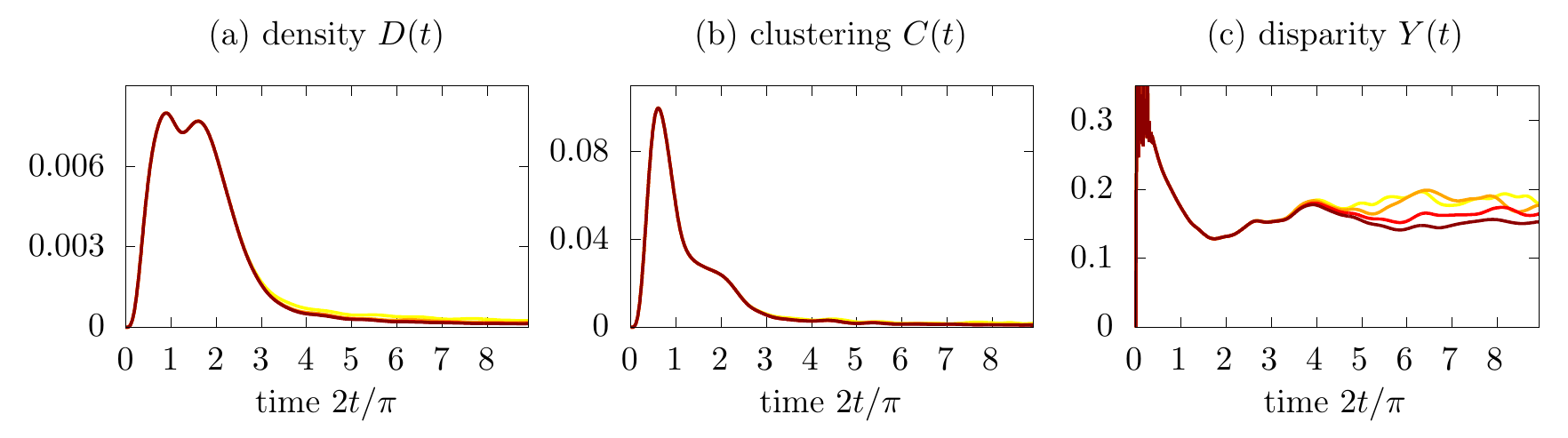}
    \caption{(color online)
    Time evolution of the network measures for the initial state $\ket{\psi(0)}=\ket{0}^{\otimes 4}\otimes\ket{1}^{\otimes 24}\otimes\ket{0}^{\otimes 4}$. The subplots show (a)  the density $D(t)$, Eq. \eqref{eq:midensity}, (b) the clustering $C(t)$, Eq. \eqref{eq:miclustering}, and (c) the disparity $Y(t)$, Eq.  \eqref{eq:midisparity}. The parameters and legenda are the same as in Fig. \ref{fig:cluster:entropy} for $L=32$. }
    \label{fig:cluster:network}
\end{figure*}

The density and clustering measures, Figs. \ref{fig:cluster:network}(a) and (b), exhibit a fast increase of correlations, which then quickly decay. After about 6 time steps, both density and clustering seem to have reached a stationary value which is small but finite. The behaviour of the disparity, subplot (d), shows that after an initial transient correlations uniformly spread across the chain.

\subsection{Evolution of a random Fock state}

\begin{figure*}
    \includegraphics[width=17cm]{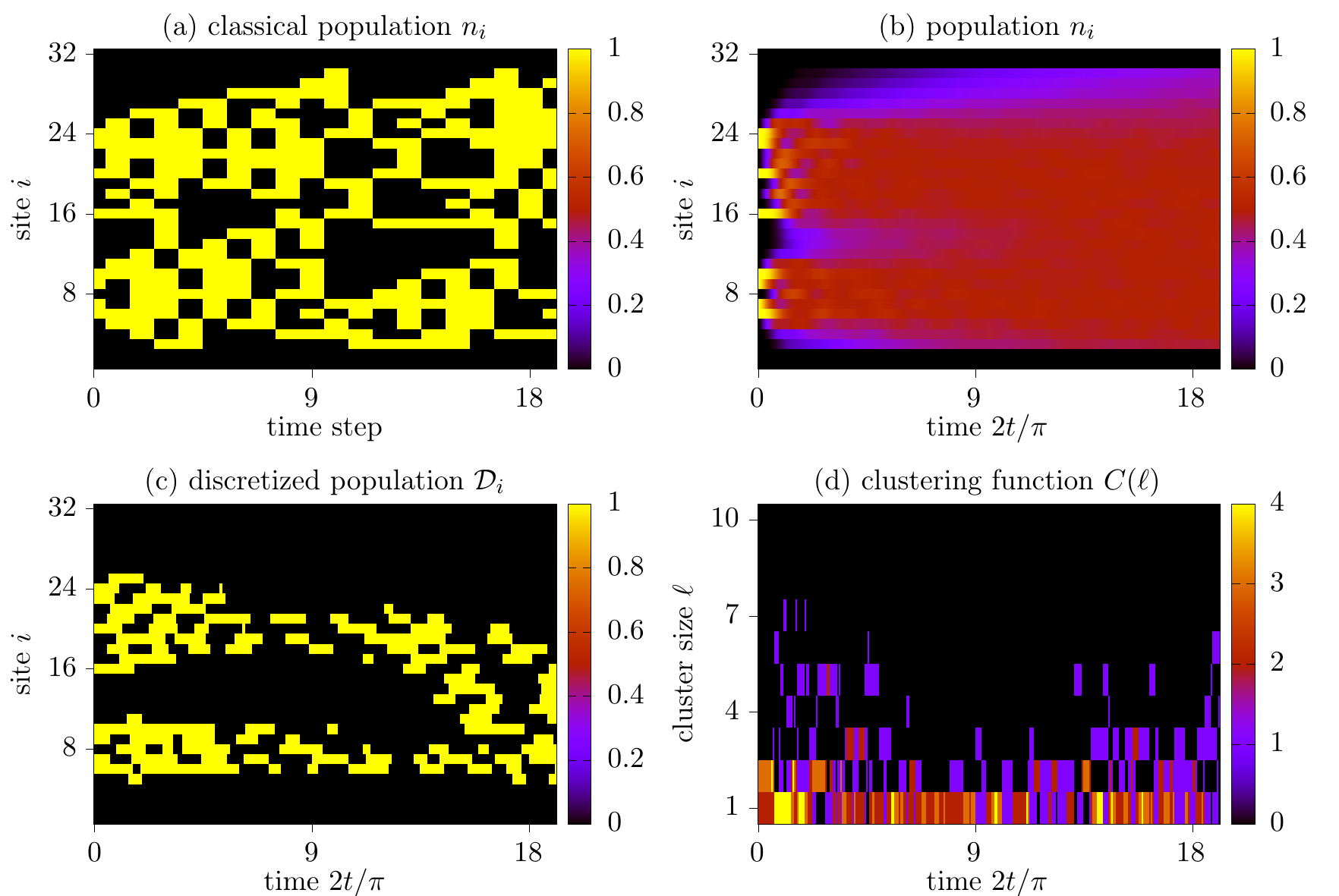}
    \caption{(color online) Characterization of the time evolution of an initial random Fock state with density $\rho(0)=0.25$. The subplots and parameters are 
the same as Fig. \ref{fig:1}, except that here for the TVDP simulations   
the bond dimension is  $m=64$ and the time step is $\Delta t=0.1$.} 
    \label{Fig:Fock}
\end{figure*}

The states we discussed so far were characterized by a symmetry in the density distribution about the center of the lattice. We now study the evolution of an initial state which is an arbitrary sequence of $0$ and $1$ states along the lattice, and which we denote by random Fock state. Figure \ref{Fig:Fock} displays an example of the dynamics of the density distribution for one specific realisation. The comparison between the classical and the quantum dynamics exhibits qualitative differences. The quantum dynamics leads to a uniform average density across the density accompanied by fluctuations in the clusters dynamics. The average size of the clusters is visibly smaller than in the classical case.

\begin{figure*}
\includegraphics[width=17cm]{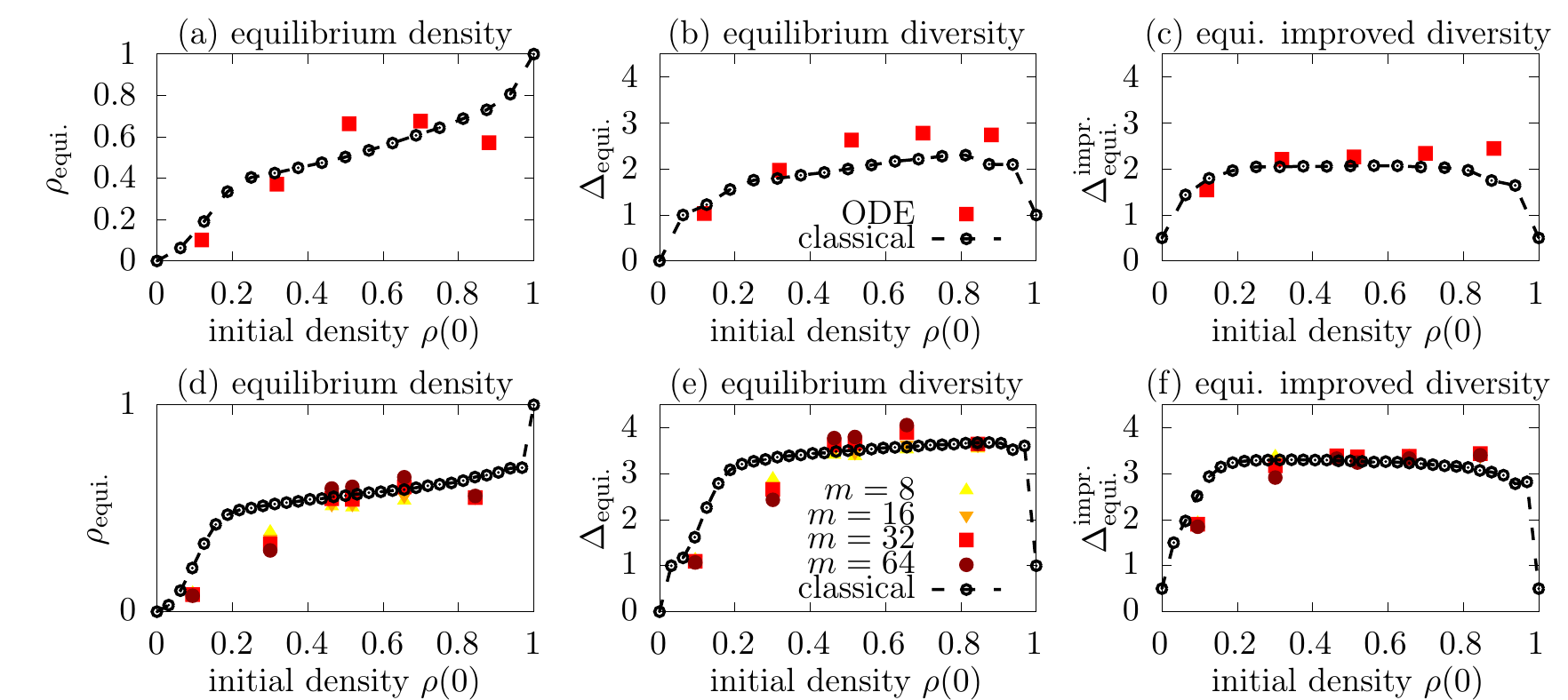}
    \caption{(color online) Characterization of the asymptotic dynamics of a random Fock 
state as a function of the initial density $\rho(0)$ and for a chain with 
$L=16$ (top) and $L=32$ sites (bottom). 
Subplots (a,d) display the equilibrium density $\rho_{\text{equi.}}$ (Eq. 
\eqref{eq:density}), (b,e) display the equilibrium diversity $\Delta_{\text{equi.}}$ (Eq. \eqref{eq:diversity}), (c,f) the equilibrium improved diversity $\Delta^{\text{impr.}}_{\text{equi.}}$, Eq. \eqref{eq:idiversity}). The black symbols indicate the results of the classical dynamics, the coloured symbols the results of TDVP simulations with bond dimensions $m=8,16,32,64$ {(darker colors correspond to higher bond dimension values, see legenda in subplot (e))} and time step $\Delta t=0.1$ (see legenda in subplot (b)). The equilibrium value is extracted by taking the time average over the time interval where the numerical results seem to have reached a stationary value. In the quantum case the interval is $[25,30]$, for the classical evolution it is  $[83,100]$.  Top row: The orange symbols are the results of the quantum dynamics integrated with a standard ODE algorithm with time step $\Delta t=0.01$. In the classical simulation, we analysed all $65\,536$ different initial configurations. Bottom row:  Each point of the classical simulation is obtained by averaging over about $5000$ initial states. In the quantum case, each point is determined by evolving 
$32$ different initial Fock states (we took the same set of states when varying the bond dimension $m$).}
    \label{fig:rd}
\end{figure*}

The long-time behavior is studied by averaging over the dynamics for different initial states with the same initial density $\rho(0)$ of alive cells. At sufficiently long times the density of alive cells, Eq. \eqref{eq:density}, the diversity, Eq. \eqref{eq:diversity}, and the improved diversity, Eq. \eqref{eq:idiversity}, tend to a constant value. In Fig. \ref{fig:rd} we compare the stationary behaviour for the classical and quantum models as a function of the initial density of the initial Fock states.
We observe interesting qualitative differences. The classical dynamics predicts that the final density is a monotonic function of $\rho(0)$ and exceeds $\rho(0)$ for $\rho(0)\lesssim 0.6$.  In the quantum dynamics, on the contrary, the final density tends to decrease again for $\rho(0)\gtrsim 0.7$. Moreover, the rate of growth at small initial densities is smaller than the classical one. The final density is maximal for $\rho(0)$ in the interval about $[0.6,0.7]$ and exceeds the classical value: quantum dynamics here leads to a larger number of alive cells than the classical game of life. The diversity shows that the quantum structures at large initial densities are more complex than the classical counterparts. The quantum improved diversity, in particular, increases monotonically with the initial density and behaves essentially as the quantum diversity, indicating that the clusters of dead cells are sparse (and of small size). The results obtained taking a larger lattice (bottom row), are consistent with these observations.

\section{Semiclassical Model for classical recurrent configurations}
\label{Sec:semi}

The appearance of quasi-recurrent configurations with the 
features of the blinkers of the classical GOL is one remarkable feature of the quantum GOL. In this section we show an example where quantization of a classical recurrent structure does not generate quantum recurrent structures. For this purpose we consider a semiclassical model of the classical rules for the evolution that combines a generic classically recurrent pattern with the analog unitary evolution and the superposition of states. In order to identify the quantum analog of a classically recurrent pattern, let us first recall the classical $F_{12}$-rule: the state of a given site changes if and only if  two or three sites among the nearest and next-nearest neighbours are alive. In a chain of length $L$, there are $2^L$ possible classical configurations. These configurations can be classified in terms of their \emph{connectivity} to other configurations. 

We now loosely employ Fock states in order to denote a sequence of 0 and 1 across the lattice. Let $\ket{\psi_0}$ be a Fock state belonging to a classical recurring configuration with periodicity $n$, and be $\ket{\psi_1},\ldots,\ket{\psi_{n-1}}$ the sequence of states obtained by applying the classical $F_{12}$ rule, such that $\ket{\psi_{n}}=\ket{\psi_{0}}$ and $n>1$.  Starting from this sequence, we now construct a corresponding quantum dynamics assuming that it solely couples the states of this sequence. We note that such dynamics will be implemented by a specific Hamiltonian, which in the subspace spanned by the states $\ket{\psi_1},\ldots,\ket{\psi_{n-1}}$ shall take the form
\begin{align}
\label{eq:newh}
\hat{H}^{(j)} &= J\sum_{k=0}^{n-1} |\psi_{k}^{(j)}\rangle\langle\psi_{k+1}^{(j)}| + |\psi_{k+1}^{(j)}\rangle\langle\psi_{k}^{(j)}|\,,
\end{align}
where $|\psi_n^{(j)}\rangle\coloneqq |\psi_0^{(j)}\rangle$.
This is equivalent to the dynamics of a quantum particle in a ring lattice of $n$ sites, where each site is one ``classical'' configuration of the 
recurrent sequence. The eigenstates can be found by using the properties of the circulant matrices, which we review in the appendix, and take the form
\begin{equation}
|\tilde\psi_\ell\rangle=\sum_{k=0}^{n-1}{\rm e}^{2\pi \mathrm{i}k\ell/n}\ket{\psi_k}/\sqrt{n}
\end{equation}
with eigenvalue 
\begin{equation*}
E_\ell=2J\cos\varphi_\ell=2J\cos(2\pi\ell/n)\,.
\end{equation*} 
{Note that these states can be formally derived as the discrete Fourier Transform of states localized in space.}
Periodic dynamics can be solely realised by superpositions 
of eigenstates whose energy differences are commensurate, which cannot be realised by means of this construction. This necessarily implies that the quantum dynamics we constructed from a classically recurrent structure does not exhibit periodicity in the spatial density distribution.

\section{Conclusions}
\label{Sec:conclusions}

The dynamics of a spin lattice, that realises a quantum analog of Conway's game of life, exhibits qualitative differences from the corresponding classical version, even when the initial state is equivalent to a classical 
string of bits. These differences are already visible in 
the dynamics of local observables. One striking feature is the 
appearance of metastable structures, which are reminiscent of classical breathers but have no classical analogon. 

{
We identify several interesting questions that will be object of future investigations. First, the study of the single-site entropy and the concurrence suggest a highly nontrivial structure of the entanglement. 
Moreover we observe that the dynamics tend to establish entanglement across the lattice following a volume-law scaling, for which it would be interesting to investigate whether the predictions of the so-called page curve are satisfied \cite{Fujita_2018,Bianchi_2019}.}
This property is a signature of the emerging complexity of 
the quantum game of life, which calls for further studies with larger lattice sizes and for its experimental implementations \cite{Whittlock2020, Farrelly2020reviewofquantum}.

\acknowledgments
We thank Frederic Folz, Rebecca Kraus, Ludger Santen, Tom Schmit and Pietro Silvi for stimulating discussions and helpful comments. 
SM acknowledges the late David Vargas, a brilliant
master's student he had the pleasure to collaborate with during his
stay at Colorado School of Mines, for insightful discussions that
partially inspired this work.
This work was funded by the Deutsche Forschungsgemeinschaft (DFG, German Research Foundation) Project-ID 429529648 TRR 306 QuCoLiMa ("Quantum Cooperativity of Light and Matter") and by the Priority Program SPP 1929 "GiRyd" (Giant Interactions in Rydberg Systems). We acknowledge the support of the German Ministry of Education and Research (BMBF) via the QuantERA projects NAQUAS and QTFLAG. 
We acknowledge support from the EU Horizon 2020 program via the Quantum Flagship
PASQUANS project, and the Italian PRIN 2017. Project NAQUAS and QTFLAG have received
funding from the QuantERA ERA-NET Cofund in Quantum Technologies implemented within the
European Union’s Horizon 2020 program.

\begin{appendix}

\section{Eigenvectors of a circulant matrix}

The matrix corresponding to Hamiltonian \eqref{eq:newh} is \emph{circulant} \cite{gray}. Consider a generic circulant matrix of size $n\times n$:
\begin{align}
H &= \mqty(
c_0 & c_1 & c_2 & \cdots & c_{n-1}\\
c_{n-1} & c_0 & c_1 & \cdots & c_{n-2}\\
c_{n-2} & c_{n-1} & c_0 & \cdots & c_{n-3}\\
\ddots & \ddots & \ddots & \ddots & \ddots\\
c_1 & c_2 & c_3 &\cdots& c_{0}
).
\end{align}
with $c_j$ is real, in our case $c_i = \delta_{i,1}+\delta_{i,n-1}$. Then the eigenvalue problem reads 
\begin{align}
\label{eq:evp}
\sum_{k=0}^{m-1} c_{n+k-m} y_k + \sum_{k=m}^{n-1} c_{k-m}y_k &= E y_m 
\end{align}
for $m=0,1,\ldots,n-1$, where $E\in\mathbb{C}$ is the eigenvalue, which is real since $H$ is hermitian, and $y_m$ are the components of the corresponding eigenvector $\vec{y} = \qty(y_0,y_1,\ldots,y_{n-1})^\intercal$. 

We make the ansatz $y_k=\rho^k$ and obtain
\begin{align}
\sum_{\alpha=n-m}^{n-1} c_{\alpha} \rho^{\alpha-n} + \sum_{\alpha=0}^{n-1-m} c_{\alpha}\rho^{\alpha} &= E.
\end{align}
where we divided both sides by $\rho^m$ and performed an index shift.
We can choose $\rho^{-n} = 1 \quad\Leftrightarrow\quad \rho_m = \exp(-2\pi \imath \frac{m}{n})$ with $m\in\mathbb{Z}$ such that $\sum_{\alpha=0}^{n-1} c_{\alpha} \rho^{\alpha} = E$. The corresponding eigenvector takes the form 
\begin{align}
\qty(\rho^0,\rho^1,\ldots,\rho^{n-1})^\intercal\,,
\end{align}
as one can prove by inserting the eigenvalue-eigenvector pair in Eq. \eqref{eq:evp}:
\begin{align}
\sum_{\alpha=0}^{n-1} c_\alpha \rho^{\alpha+m}
&= \sum_{\alpha=0}^{n-1} c_\alpha \rho^\alpha \rho_m\,. 
\end{align}
Here, we used $\rho^m=\rho^{m+n}=\rho^{m-n}$ for $m=0,1,\ldots,n-1$.
Inserting the values of $c_j$ we obtain
\begin{align}
E_m &= \sum_{\alpha=0}^{n-1} c_\alpha \rho^{\alpha}_m
%&= \sum_{\alpha=0}^{n-1} c_\alpha \exp(-2\pi\imath \frac{m\alpha}{n})\\
%&= \exp(-2\pi\imath \frac{m(1)}{n}) + \exp(-2\pi\imath \frac{m(n-1)}{n})\\
%&= \exp(-2\pi\imath \frac{m(1)}{n}) + \exp(-2\pi\imath \frac{m(-1)}{n})\\
= 2\cos(2\pi \frac{m}{n})\,,
\end{align}
with the corresponding eigenvector
\begin{align}
\vec{y}_m &= \mqty(1\\ \exp(-2\pi\imath \frac{m}{n})\\ \exp(-2\pi\imath 
\frac{2m}{n})\\\vdots\\\exp(-2\pi\imath \frac{(n-1)m}{n}))\eqqcolon \sqrt{n}\ket{y^m}.
\end{align}

\end{appendix}

\end{document}